\documentstyle[preprint,prl,aps]{revtex}
\tighten
\begin{document}
\draft
\preprint{}
\title{ Non-equilibrium Dynamics Following a Quench to the Critical Point in
a Semi-infinite System}

\author{Satya N. Majumdar}
\address{Yale University\\
Department of Physics,  Sloane Physics Laboratory\\
New Haven, CT-06520-8120; USA.\\
e-mail: satya@cmphys.eng.yale.edu}

\author{Anirvan M. Sengupta}
\address{AT\&T Bell Laboratories\\
600 Mountain Avenue, Murray Hill, NJ 07974; USA.\\
e-mail: anirvan@physics.att.com}

\maketitle
\begin{abstract}
We study the non-equilibrium dynamics (purely dissipative and relaxational)
in a semi-infinite system following a quench from the high temperature
disordered phase to its critical temperature. We show that the local
autocorrelation near the surface of a semi-infinite system decays
algebraically in time
with a new exponent which is different from the bulk. We calculate this new
non-equilibrium surface exponent in several cases, both analytically and
numerically.
\end{abstract}
\vspace{2cm}


\narrowtext
\newpage
There has been a lot of current interest in understanding the growth of
correlations in a system after being quenched from the high temperature
disordered phase to or below its critical temperature ($T_c$) \cite{AB}. In
either case
the system exhibits dynamic scaling at late stage of the growth. The
growth is characterized by a single time dependent length scale. For quench to
below $T_c$, this length scale characterizes the linear size of the growing
domains of competing broken symmetry phases. On the other hand, for quench to
$T_c$, it characterizes the length scale over which equilibrium critical
properties are established. A lot of theoretical and experimental efforts have
been directed in determining
the time dependence of this length scale and the scaling of the equal-time
correlation functions. It was, however, realized later that even the
two-time correlation functions have interesting dynamical scaling \cite{FH,NB}.
In particular, the auto-correlation function, measuring the memory of the
initial conditions retained by the system after time $t$, decays algebraically
with time \cite{FH}. This has been verified by exact calculations in a few
cases \cite{AB2,AF,CZ,MHL,BD,RB},
numerical simulations \cite{BH} and very recently experimentally \cite{MPY} for
quench to $T<T_c$, in a liquid crystal system using video-microscopy.

In static critical phenomena, it is well known that the critical behaviour
near the boundary of a semi-infinite system is drastically different from
the behaviour deep inside the bulk \cite{BH,LR,BM,JC,IPT}. It is therefore
natural and important
to know whether there are similar modifications in the dynamical behaviour near
the boundary. For the critical dynamics following a quench to $T_c$, it has
been
argued that even the semi-infinite system is characterized by a single
time dependent length scale $\xi (t)$ which has the same growth law as
the infinite system \cite{DD,KO}. However, in this letter we demonstrate,
both analytically
and numerically, that the temporal decay of
the critical auto-correlations near the boundary of a semi-infinite
system is characterized by new exponents different from that in the bulk.

We consider a semi-infinite $O(n)$
model in the space $[\vec x=(\vec r, z)]$ which extends over infinite space in
$d-1$ directions (denoted by $\vec r$) and over
only positive half-space in one direction ($z\geq 0$). The system is
assumed to be translationally invariant in the $(d-1)$ directions and this
invariance is broken in the $z$ direction due to the presence of a surface at
$z=0$. The model is described
by an $n$ component order parameter field ${\vec \phi}
=[{\phi}_1,\ldots, {\phi}_n]$ and a coarse
grained Landau-Ginzburg free energy functional with an additional surface
contribution \cite{BM},
\begin{equation}
F(\vec \phi)={1\over {2}}\int d^d\,{\vec x} [(\nabla {\vec \phi})^2+r_0
{\vec \phi}^2+ {u\over {4}}({\vec \phi}^2)^2 + c\delta (z) {\vec \phi}^2],
\label{action}
\end{equation}
where the integration in Eq. (1) is over the half-space $z\geq 0$. The
equilibrium properties of this model have been studied in detail \cite{LR,BM}.
Depending upon the value of $c$, different types of surface orderings take
place. There
exists a special value $c=c^{*}$ such that, for $c>c^{*}$, the surface orders
along with the bulk at bulk $T_c$. This parasitical transition is called
``ordinary'' transition \cite{LR,BM}. For $c<c^{*}$ and in high enough
dimensions
(such that a $d-1$ dimensional surface can order), the
surface orders first as the temperature is lowered while the bulk is still
disordered (``surface'' transition) and then as the temperature is lowered
further the bulk orders in presence of an ordered surface (``extraordinary''
transition). The value $c=c^{*}$ is a special point where the critical
exponents are different from the ordinary or surface transitions. This
is called the ``special'' transition. Within mean field theory, $c^{*}=0$ but
becomes nonzero for $d<4$ due to corrections arising from fluctuations
\cite{BM}. The critical
exponents associated with these different types of transitions are different
from each other and from the bulk values \cite{BM}. For example, an exact
calculation
exists \cite{MW} for the $2$-d semi-infinite Ising model
(where the surface undergoes an ``ordinary''
transition at the bulk $T_c$ \cite{BM}) which shows that the
two-point correlator between two-points on the surface (separated by $r$)
decays faster as $r^{-1}$ for large $r$ than the bulk correlator that
decays as $r^{-1/4}$.

In this paper, we consider the non-conserved dynamics of the order parameter
field in presence of a surface following a quench from the high temperature
disordered $(T>T_c)$ to the bulk critical point $T=T_c$ and ask whether the
presence of the surface modifies the dynamics near the surface. Far from
the surface one should recover the critical dynamics of a truly infinite
system for which
several results are known. For example, it is now well established
\cite{JSS,DH}that the
bulk equal-time correlation function, $G(\vec x,t)
=\langle \phi (\vec x',t) \phi (\vec x'+\vec x,t)\rangle$ exhibits dynamic
scaling, $G(\vec x,t)\sim x^{-(d-2+\eta)}g_c(x/{\xi (t)})$ where $g_c$ is a
universal scaling function and $\xi (t)\sim t^{1/{\cal Z}}$ is the
time-dependent
correlation length. $\eta$ and $\cal Z$ are the usual static and dynamic
exponents and the $\langle \rangle$
denotes an average over all possible initial conditions (corresponding to
the equlibrium distribution at the initial high temperature) and over the
history of time evolution. The bulk two-time correlation function
$C(\vec x,t)=
\langle \phi (\vec x',0) \phi (\vec x'+\vec x, t)\rangle$, measuring the
correlation with the initial condition, also exhibits dynamic scaling
\cite{JSS,DH},
$C(\vec x,t)\sim [\xi (t)]^{-{\lambda}_c}f_c(x/{\xi (t)})$ where $f_c(0)$ is a
constant of $O(1)$. The exponent ${\lambda}_c$, characterizing the
decay of the bulk autocorrelation, $A_b(t)=\langle {\phi} (\vec x,0) {\phi}
(\vec x,t)
\rangle \sim [\xi (t)]^{-{\lambda}_c}$, is a new critical exponent
\cite{DH,JSS}
in the sense that no simple scaling relation has been found relating it to
other static or dynamic critical exponents. For an infinite system,
${\lambda}_c$ has been calculated analytically for the $O(n)$ model in the
limit $n\to \infty$ and also within $\epsilon$-expansion where
${\epsilon}=4-d$ ($d=4$ being the upper critical dimension) \cite{JSS}. For
Ising model
in $d=2$ and $3$, ${\lambda}_c$ has been determined numerically \cite{DH}.

The specific dynamical quantity that we calculate explicitly in this paper for
the semi-infinite system and
show that it gets drastically modified due to the presence of the surface, is
the decay
of the auto-correlation $A(z, t)=\langle {\phi}(\vec r, z, 0)
{\phi}(\vec r, z,t)\rangle$ with time $t$.
In the limit $z\to \infty$, we recover, as expected,
the bulk results $A(\infty ,t)\sim [\xi (t)]^{-{\lambda}_b}$ where we
denote the bulk ${\lambda}_c$ by ${\lambda}_b$. However, for small $z$ near
the surface, we find that the autocorrelator decays as $A_s(t)\sim
[\xi (t)]^{-{\lambda}_s}$ where ${\lambda}_s$ is a new dynamical surface
exponent different from ${\lambda}_b$. Also, the value of ${\lambda}_s$ depends
explicitly on the type of the surface transition. In this paper, we calculate
${\lambda}_s$ analytically within $\epsilon$-expansion and in the $n\to \infty$
limit for the ``ordinary'' and ``special'' transition. Also, we determine the
value of ${\lambda}_s$ numerically for the two-dimensional Ising model.

The model-A dynamics of the order parameter is governed by the Langevin
equation,
\begin{equation}
{{\partial {\vec \phi}}\over {\partial t}}=-{{\delta F}\over {\delta
{\vec \phi}}}+ {\vec \eta}
\label{langevin}
\end{equation}
where $F$ is given by Eq. [1] and ${\vec \eta} (\vec x,t)$ is a Gaussian noise
with zero average and a correlator $\langle {\eta}_i (\vec x,t) {\eta}_j
(\vec x',t')\rangle= 2 k_BT {\delta}_{i,j} \delta(\vec x -\vec x')\delta
(t-t')$, where $T$ is the temperature. We first consider the Gaussian theory
where one neglects the interaction (set $u=0$ in Eq. (1)) and which is valid
for $d>4$. We define the Fourier transform, $G(\vec k, z, z',t)=\int d^{d-1}
(\vec r-\vec r') G(\vec r -\vec r', z, z', t) \exp [i{\vec k}.(\vec r-\vec r')]
$ where $\vec k$ is a $(d-1)$ dimensional vector in the reciprocal
space. Then from Eq. (2), at the critical point
($r_0=0$ and setting  $k_B T_c=1$), $G(\vec k, z,z',t)$ evolves as,
\begin{equation}
{\partial}_t G(\vec k, z,z',t)=[-2k^2 + {{\partial}_z}^2 +{{\partial}_{z'}}^2]
G(\vec k, z,z',t) + 2 \delta (z-z')
\label{eqtime}
\end{equation}
with the boundary condition, ${{\partial}_z}G=c G$ at $z=0$ and the initial
condition, $G(\vec k, z,z',0)=\Delta \delta (z-z')$ (this "white noise" form
of the initial condition corresponds to
quench from the infinite temperature where the field $\phi (\vec x, t)$ is
completely random and $\Delta$ controls the size of initial onsite fluctuations
in $\phi$). Similarly, the symmetrized two-time correlation
function $C_s(\vec k, z,z', t)$ defined as the Fourier transform of
$C_s(\vec r, z, z', t)={1\over 2}\langle ({\phi}(\vec r', z',0){\phi}(\vec r'
+\vec r, z, t) +{\phi}(\vec r', z,0){\phi}(\vec r'+\vec r, z', t)
\rangle$ evolves as
\begin{equation}
{\partial}_t C_s(\vec k, z,z',t)={1\over 2}[-2k^2 + {{\partial}_z}^2
+{{\partial}_{z'}}^2]C_s(\vec k, z,z',t)
\label{twotime}
\end{equation}
with the same boundary and initial conditions. By choosing the basis function
\begin{equation}
\psi (u,z)={1\over {\sqrt 2}}[\exp (iuz)-{{c-iu}\over {c+iu}}\exp (-iuz)],
\label{eigen}
\end{equation}
it is easy to see that the solutions to Eq. (3) and (4) are given by,
$G(\vec k, z,z',t)=\int_{-\infty}^{\infty}du {\psi}(u,z){\psi}^{*}(u, z')
f_1(k,u,t)$ and $C_s(\vec k, z, z', t)=\int_{-\infty}^{\infty}du {\psi}(u,z)
{\psi}^{*}(u, z') f_2(k,u,t)$ where
\begin{equation}
f_1(k,u,t)=\Delta \exp [-2(k^2+u^2)t] +{{1-\exp [-2(k^2+u^2)t]}\over {k^2+u^2}}
\label{f1}
\end{equation}
and
\begin{equation}
f_2(k,u,t)=\Delta \exp [-(k^2+u^2)t].
\label{f2}
\end{equation}
Therefore, the autocorrelation $A(z,t)=\int C_s(\vec k, z,z,t){{d^{d-1}\vec k}
/ {(2\pi)^{d-1}}}$ is given by
\begin{equation}
A(z,t)\sim t^{-(d-1)/2}\int_0^{\infty}du \exp (-u^2t){{(c\sin {uz}+ u\cos
{uz})^2}\over {c^2+u^2}}.
\label{auto}
\end{equation}
It is clear from Eq. (8) that in the limit $z\to \infty$, we recover the
bulk result: for large $t$, $A(\infty, t)\sim [\xi (t)]^{-d}$ where
$\xi (t)\sim t^{1/2}$
(${\cal Z}=2$ within Gaussian theory) and hence ${\lambda}_b=d$. On the other
hand,
for $z=0$, we find that for large $t$, $A(0,t)\sim [\xi (t)]^{-(d+2)}$ for
$c>0$ and $A(0,t)\sim [\xi (t)]^{-d}$ for $c=0$. Thus, we obtain the
results that for the ``special'' transition ($c=0$) ${\lambda}_{sp}=d$ while
for the ``ordinary'' transition ($c>0$), ${\lambda}_{or}=d+2$ within the
 Gaussian theory. It is interesting to note that while ${\lambda}_b$
satisfies the
upper bound ${\lambda}_b\leq d$ conjectured by Fisher and Huse \cite{FH},
clearly ${\lambda}_{or}$ violates this upper bound.

For $d<4$, where the interaction term is no longer irrelevant, we evaluate the
exponent ${\lambda}_c$ in $\epsilon=4-d$ expansion. The two-time
correlator (unsymmetrized) $C(\vec x, \vec x', t)=\langle \phi (\vec x',0)
\phi (\vec x, t)\rangle$ in real space evolves as
\begin{equation}
{\partial}_t C(\vec x, \vec x', t)=[-r_0+{\nabla}^2] C(\vec x, \vec x', t)
-{{u_0}\over {n}}\sum_{ij}\langle {\phi}_i(\vec x',0) {\phi}_i(\vec x,t)
{\phi}_j(\vec x,t) {\phi}_j(\vec x,t)\rangle.
\label{evolve}
\end{equation}
At the Wilson-Fisher fixed point, $u_0={{8{\pi}^2}\epsilon}/(n+8)$
to leading order in $\epsilon$ \cite{BM}. This allows one to calculate the
corrections to the two-point correlator perturbatively in $u_0$. To leading
order in $\epsilon$, the term proportional to $u_0$ in Eq. (9) can be
expressed, using Wick's theorem, in terms
of the mean field propagators as $-u_0(n+2)G_0(\vec x, \vec x, t)C_0(\vec x',
\vec x, t)$ where $G_0$ and $C_0$ denote the mean field equal-time and
two-time propagators respectively. To
leading order in $\epsilon$, one can replace $C_0$ in this term by $C$ and
then Eq.(9) becomes a linear evolution equation for the two-time correlator
which is correct to $O(\epsilon)$. This evolution equation, for the symmetrized
correlator $C_s(\vec x, \vec x', t)$, reads
\begin{equation}
{\partial}_t C_s(\vec x, \vec x', t)={1\over 2}[-2{\tilde r_0}+
{\nabla_{\vec x}}^2+{\nabla_{\vec x'}}^2 -
V(z,t)-V(z',t)] C_s(\vec x,\vec x', t)
\label{lineq}
\end{equation}
with ${\tilde r_0}=r_0+u_0(n+2)G_0(\infty, \infty, \infty)$ where $G_0 (\vec x,
\vec x', t)$ denotes the mean-field equal-time propagator. The potential
$V(z,t)=u_0(n+2)\int  [G_0(\vec k, z,z,t)-
G_0(\vec k, \infty, \infty, \infty)]{{d^3(\vec k)}/ {(2\pi)^3}}$ captures
the corrections due to fluctuations
for $d<4$ and can be calculated explicitly. For example, at the ``special''
($c=0$) and the ``ordinary'' ($c=\infty$) fixed points, we get, $V_{sp}(z,z,t)
=-{u_0(n+2)[1+(2t/z^2)\exp (-z^2/{2t})]}/{32{\pi}^2 t}$ and
$V_{or}(z,z,t)
=-{u_0(n+2)[1-(2t/z^2)\exp (-z^2/{2t})]}/{32{\pi}^2 t}$ in
the scaling regime where $z>>{\Lambda}^{-1}$, ${\Lambda}$ being the upper
cutoff.

The Eq. (10) and the form of the potential $V(z,t)$ suggest a
late time scaling ansatz for
the Fourier transform $C_s(\vec k, z,z',t)\approx t^{-\alpha} \exp (-k^2 t)
f[z/{\sqrt t}, z'/{\sqrt t}]$. To determine the exponent $\alpha$ we first
consider the bulk limit $z\to \infty$, $z'\to \infty$.  Using $V(\infty,
\infty,t)=-u_0(n+2)/{32 {\pi}^2 t}$  and $u_0 = {{8\pi^2 \epsilon}/ (n+8)}$
in Eq. (10) we get $\alpha={1 \over 2} - {(n+2) \epsilon}/{4(n+8)}$
and $f(x,y)\sim e^{-(x-y)^2}$ as $x,y \to \infty$. It follows
immediately that the bulk autocorrelation $A_b (t) \sim t^{- (d  -
{{n+2 }\over {n+8}} {\epsilon \over 2})/2}$. Since the dynamic exponent
${\cal Z}= 2+O({\epsilon} ^2)$, we recover the bulk result $\lambda_b= d -
{{n+2 }\over {n+8}} {\epsilon \over 2}$. For the surface autocorrelator,
we need to know the small argument behaviour of the scaling function
$f(x,y)$. For small $x,y$, $f(x,y)\sim (xy)^s (a+bx^2 +c y^2 +\cdots)$
where $s(s-1)=\pm {{n+2 }\over{n+8}} {\epsilon \over 2}$ and $\pm$
corresponds to ``special'' and to ``ordinary'' transitions respectively. Then
the autocorrelator $A(z,t) =\int C_s(\vec k,z,z,t){d^{d-1}k / {(2\pi)
^{d-1}}} \sim t^{-(d-1+2s+2\alpha)/2}$. We choose the root of $s$ to match the
$\epsilon \to 0$ limit and get  $\lambda_{sp}=d-  {{n+2 }\over{n+8}} {3\epsilon
 \over 2}$ for ``special'' transition and  $\lambda_{or} =d+2-{{n+2
}\over{n+8}}
{3\epsilon \over 2}$ for the ``ordinary'' one.

We next calculate ${\lambda}_c$ exactly in the large $n$ limit. In this limit,
$C_s(\vec x, \vec x',t)$  satisfies Eq. (10) exactly except that
the potential $V(z,t)$ is determined self-consistently from $V(z,t)=
u_0(n+2)\int  [G(\vec k, z,z,t)-
G(\vec k,\infty, \infty, \infty)]{{d^{d-1}(\vec k)}/ {(2\pi)^{d-1}}}$, where
the equal-time propagator $G(\vec k,z,z',t)$ satisfies:
\begin{equation}
{\partial}_t G(\vec k, z,z',t)=[-2k^2 + {{\partial}_z}^2 +{{\partial}_{z'}}^
2-V(z,t)-V(z',t)]
G(\vec k, z,z',t) + 2 \delta (z-z').
\label{glargen}
\end{equation}
In analogy with epsilon expansion, we make the ansatz $V(z,t)=
{a\over {2t}} +{{(\mu^2 -{1/ 4})} \over z^2} g[z/{\sqrt t}]$ where
$g(x)\to 1$ as $x\to 0$ and $g(x)\to 0$ as $x\to \infty$. The values
of $a$ and  $\mu$ are determined respectively from the
limits $z\to \infty$ (bulk dynamics) and $t\to \infty$ (static limit)
and are already known to be $a={(d-4)/ 2}$ \cite{JSS}, $\mu_{sp}=
{(d-5)/ 2}$
and $\mu_{or}={(d-3)/ 2}$ \cite{BM}. The full form of the scaling function
$g(x)$
is to be determined from a complicated self-consistent equation. However,
for the purpose of determining the surface exponent $\lambda_s$,
it is sufficient to know that $g(0)=1$. We then proceed identically
as in the case of $\epsilon $ expansion by assuming a scaling ansatz
for $C_s(\vec k,z,z',t)$. We find $\alpha={(1+a)/2}$ and
$s(s-1)=\mu^2- {1/4}$. Using the fact that ${\cal Z}=2$ in the large $n$
limit, we obtain $\lambda_{sp}={(5d-12)/ 2}$  and
 $\lambda_{or}={(5d-8) / 2}$. Note that in the limit $z\to \infty$
we recover the bulk result $\lambda_b= {(3d-4)/2}$. These results are
consistent  with those obtained from $\epsilon$ expansion after taking
$n\to \infty$ limit and also with the mean field results in $d=4$.

We have also carried out a direct numerical simulation of the two-dimensional
Ising model with open boundary conditions at the bulk critical temperature
using spin-flip Metropolis algorithm. We measure  boundary spin correlations
and compare them with the corresponding bulk measurements done with periodic
boundary conditions. In the static limit, it is well known
that the boundary spins order only due to the ordering of the
bulk spins (``ordinary'' transition) \cite{BH,LR,BM}. Since at $T_c$ in the
static limit, the
boundary correlator falls off with distance $r$ as $1/r$ (as opposed to
the bulk decay  $r^{-1/4}$) for large $r$ \cite{MW},  it is natural to assume
a dynamic scaling form for the equal-time boundary correlator
$G_s(r,t)\sim r^{-1} \gamma (r/{\xi (t)})$ at late times. It is believed
that even in the presence of a boundary, there is still only a single
time-dependent correlation length $\xi (t) \sim
t^{1/{\cal Z}}$ with ${\cal Z}\approx 2.15$ numerically \cite{DD,KO}. The
quantity
$\chi_s(t)=\langle (M_s(t))^2 \rangle $ ($M_s(t)$ being the total boundary
magnetisation at time $t$) is the  integral of the correlation function
$\int_s dr G_s(r,t)$ on the boundary. From the scaling form of $G_s(r,t)$,
$\chi_s(t)$ would then grow logarithmically with $\xi (t)$ (and hence
logarithmically with $t$). In contrary, the corresponding
quantity in the bulk, $\chi_b(t)=\langle (M_b(t))^2\rangle $ ($M_b$ being the
total bulk magnetisation), grows algebraically as $[\xi (t)]^{7/4}$.
In Fig. 1 we plot $\chi_s(t)$ and $\chi_b(t)$ against $\log (t)$ and, in the
inset, show a log-log plot of $\chi_b(t)$ versus $t$.  Next, we compute
the auto-correlation on the boundary of the open system and the periodic bulk
system.  The two results
are contrasted in Fig. 2. The autocorrelation on the boundary decays much
faster. This is expected for ``ordinary'' transition as evident from the
large $n$ and the $\epsilon$ expansion calculations. From the slope of the
log-log plot in Fig. 2  we estimate the boundary exponent ratio
$\lambda_{or}/{\cal Z}=1.2 \pm 0.1$. The corresponding ratio in the bulk is
estimated
to be  $\lambda_b/{\cal Z}=.74 \pm 0.02$ in agreement with the previous
simulation \cite{DH}. Using ${\cal Z} \approx 2.15$, we estimate,
${\lambda}_{or}
\approx 2.58\pm 0.1$ to be contrasted with ${\lambda}_b\approx 1.59\pm 0.02$.

A two-dimensional system where these boundary auto-correlation exponents
can be calculated exactly is the dynamics of the X-Y model following a
quench from  one temperature to another, both temperatures being below the
Kosterlitz-Thouless temperature. In this case, the system is free of vortices
and therefore, the dynamics of the phases is trivially that of damped
independent spin waves. However, as shown by Rutenberg and Bray \cite{RB}, the
bulk auto-correlation decays algebraically with time and the corresponding
exponent  $\lambda_b$ depends continuously on temperature. It is
straightforward to extend their calculations to systems with edges and
corners. For example, we have found that $\lambda_{corner}=2\lambda_{edge}
=4\lambda_b$. Details of these calculations will be published elsewhere
\cite{SM}.

It is clear from the results presented in this paper that for quench
to $T=T_c$, the surface autocorrelation exponents are quite different
from their bulk counterparts. It is an interesting question to ask whether
the same happens for  quench to the ordered phase ($T<T_c$). From our
preliminary $T=0$ simulations of the $2d$ Ising model, $\lambda_s$ does
not seem to be different from $\lambda_b$ at $T=0$. This seems to be the case
also for the exact $T=0$ Glauber dynamics of the semi-infinite Ising chain.
Interpretation of these
results and further studies are relegated to a future publication \cite {SM}.

\acknowledgements
We thank David Huse and Subir Sachdev for very useful discussions. S.M's
research was funded by NSF grant no. DMR-92-24290.

\begin{figure}
\caption{ $\chi_s(t)$ is plotted  against $\log (t)$. The logarithmic
dependence is pretty evident.
The inset shows a log-log plot of $\chi_b(t)$ versus $t$, which
is consistent with a
power law.      }
\label{chi}
\end{figure}

\begin{figure}
\caption{The autocorrelation on the boundary, $A_s(t)$ versus time $t$ in a
log-log plot.
{}From the slope  we estimate the boundary exponent ratio
$\lambda_{or}/{\cal Z}=1.2 \pm 0.1$. The corresponding ratio in the bulk is
estimated
to be  $\lambda_b/{\cal Z}=.74 \pm 0.02$. Bulk data is shown in the inset. }
\label{auto}
\end{figure}


\begin{references}

\bibitem{AB} For a recent review, see A.J. Bray, {\sl Advances in Physics},
{\bf 43}, 357 (1994).

\bibitem{FH} D.S. Fisher and D.A. Huse, {\sl Phys. Rev. B}, {\bf 38}, 373
(1988).

\bibitem{NB}T.J. Newman and A.J. Bray, {\sl J. Phys. A., Math. Gen.}, {\bf 23},
4491 (1990).

\bibitem{AB2}A.J. Bray, {\sl J. Phys. A, Math. Gen. }, {\bf 22}, L67 (1990).

\bibitem{AF}J.G. Amar and F. Family, {\sl Phys. Rev. A}, {\bf 41}, 3258 (1990).

\bibitem{CZ} A. Coniglio and M. Zannetti, {\sl Europhys. Lett.}, {\bf 10},
575 (1989).

\bibitem{MHL} S.N. Majumdar, D.A. Huse, and B.D. Lubachevsky, {\sl Phys. Rev.
Lett.}, {\bf 73}, 182 (1994); S.N. Majumdar and D.A. Huse, {\sl to appear
in Phys. Rev. E.}, 1995; C. Sire and S.N. Majumdar, {\sl Phys. Rev. Lett.},
{\bf 74}, 4321 (1995).

\bibitem{BD} A.J. Bray and B. Derrida, {\sl Phys. Rev. E.}, {\bf 51},
R1633 (1995).

\bibitem{RB}A.D. Rutenberg and A.J. Bray, {\sl Phys. Rev. E.}, {\bf 51},
R1641 (1995).

\bibitem{BH} T.J. Newman, A,J. Bray, and M.A. Moore, {\sl Phys. Rev. B},
{\bf 42}, 4514 (1990); A.J. Bray  and K. Humayun, {\sl J. Phys. A, Math. Gen.},
{\bf 23}, 5897 (1990).

\bibitem{MPY}N. Mason, A.N. Pargellis, and B. Yurke, {\sl Phys. Rev. Lett.},
{\bf 70}, 190 (1993).

\bibitem{DH}D.A. Huse, {\sl Phys. Rev. B}, {\bf 40}, 304 (1989).

\bibitem{JSS}H.K. Janssen, B. Schaub, and B. Schmittman, {\sl Z. Phys. B},
{\bf 73}, 539 (1989).

\bibitem{BH}K. Binder and P.C. Hohenberg, {\sl Phys. Rev. B}, {\bf 6},
3461 (1972).

\bibitem{LR} T.C. Lubensky and M.H. Rubin, {\sl Phys. Rev. Lett.}, {\bf 31},
1469 (1973); {\sl Phys. Rev. B.}, {\bf 12}, 3885 (1975).

\bibitem{BM}A.J. Bray and M.A. Moore, {\sl J. Phys. A: Math. Gen.}, {\bf 10},
1927 (1977).

\bibitem{JC} J.L. Cardy, {\sl Phase Trasitions and Critical Phenomena, vol. 11,
edited by C. Domb and J.L. Lebowitz (London, Academic}, p55 (1987).

\bibitem{IPT}F. Igloi, I. Peschel, and L. Turban, {\sl Advances in Physics},
{\bf 42}, 683 (1993).

\bibitem{MW} B.M. Mccoy, and T.T. Wu, {\sl Phys. Rev.}, {\bf 162}, 436 (1967).

\bibitem{DD}S. Dietrich and H.W. Diehl, {\sl Z. Phys. B}, {\bf 51}, 353 (1983).

\bibitem{KO}M. Kikuchi and Y. Okabe, {\sl Phys. Rev. Lett.}, {\bf 55}, 1220
(1985); H. Riecke, S. Dietrich, and H. Wagner, {\sl Phys. Rev. Lett.}, {\bf
55}, 3010 (1985); H. W. Diehl, {\sl Phys. Rev. B}, {\bf 44}, 2846 (1994).

\bibitem{SM}A.M. Sengupta and S.N. Majumdar, {\sl in preparation}.

\end{references}
\end{document}